# The Diversity of Gamification Evaluation in the Software Engineering Education and Industry: Trends, Comparisons and Gaps

Rodrigo Henrique Barbosa Monteiro
*Graduate Program in Computer Science*
*Federal University of Pará*
Belém, Pará, Brazil
rodrigo.monteiro@icen.ufpa.br

Maurício Ronny de Almeida Souza
*Graduate Program in Computer Science*
*Federal University of Lavras*
Lavras, Minas Gerais, Brazil
mauricio.ronny@ufla.br

Sandro Ronaldo Bezerra Oliveira
*Graduate Program in Computer Science*
*Federal University of Pará*
Belém, Pará, Brazil
srbo@ufpa.br

Carlos dos Santos Portela
*Faculty of Information Systems*
*Federal University of Pará*
Belém, Pará, Brazil
csp@ufpa.br

Cesar Elias de Cristo Lobato
*Faculty of Information Systems*
*Federal University of Pará*
Belém, Pará, Brazil
cesarlobato27@gmail.com

*Abstract*— **Context**: gamification has been used to motivate and engage participants in software engineering education and practice activities. **Problem**: There is a significant demand for empirical studies for the understanding of the impacts and efficacy of gamification. However, the lack of standard procedures and models for the evaluation of gamification is a challenge for the design, comparison, and report of results related to the assessment of gamification approaches and its effects. **Goal**: The goal of this study is to identify models and strategies for the evaluation of gamification reported in the literature. **Method**: To achieve this goal, we conducted a systematic mapping study to investigate strategies for the evaluation of gamification in the context of software engineering. We selected 100 primary studies on gamification in software engineering (from 2011 to 2020). We categorized the studies regarding the presence of evaluation procedures or models for the evaluation of gamification, the purpose of the evaluation, the criteria used, the type of data, instruments, and procedures for data analysis. **Results**: Our results show that 64 studies report procedures for the evaluation of gamification. However, only three studies actually propose evaluation models for gamification. We observed that the evaluation of gamification focuses on two aspects: the evaluation of the gamification strategy itself, related to the user experience and perceptions; and the evaluation of the outcomes and effects of gamification on its users and context. The most recurring criteria for the evaluation are "engagement", "performance", "satisfaction", and "motivation". Finally, the evaluation of gamification requires a mix of subjective and objective inputs, and qualitative and quantitative data analysis approaches. Depending of the focus of the evaluation (the strategy or the outcomes), there is a predominance of a type of data and analysis.

## I. Introduction

Gamification is the use of game design elements in non-game contexts [1], for the purpose of motivating and engaging users in activities present in these contexts. This approach is used in software engineering education and software development to promote behavior and psychological changes [2] that are able to improve performance, efficiency and engagement in educational and development activities.

Therefore, new studies have been published advocating and assessing these outcomes in software engineering related areas. Souza et al. [3] present a systematic mapping study on game-related approaches, and their results show gamification as a growing topic since 2011. However, Souza et al. [3] and Jiang [4] suggest the need of additional empirical data and studies to support the effects of gamification in software engineering education.

Nonetheless, few studies propose methods to evaluate gamification in software engineering [5, 6, 7]. As a consequence, each gamification study proposes its own approach to evaluate gamification, and/or identifies the key aspects that should be evaluated in its specific scope in future studies. However, these projects may take long to mature and provide significant data to support its usefulness and efficacy. As a result, secondary studies are especially important to allow the definition of gamification evaluation models, based on the state-of-the-art and state-of-the-practice of gamification in software engineering, as reported in the literature.

The goal of this paper is to identify primary studies in software engineering that evaluate or propose evaluation models for gamification. From these studies, we aim to summarize the main goals, methods, criteria and analysis procedures used in the literature for the evaluation of gamification. To achieve this goal we executed a systematic mapping study and selected 100 studies on gamification in software engineering (both education and practice) for the analysis.

In our results, 64 studies propose or applied strategies for the evaluation of gamification. However, we found only three studies that actually propose evaluation models for gamification in the context of software engineering. We identified that the evaluation of gamification focus on (i) the approach itself and the user experience, or (ii) on the outcomes and effects of gamification on its users. Finally, we observed 4 specially recuring criteria for the evaluation of gamification: engagement, performance, satisfaction and motivation. We hope that our results may support the design and evaluation of gamification, which has become gradually more present in software engineering education and training contexts.

The reminder of the paper is organized as follows. Section II provides the necessary theoretical foundation for the study. Section III describes the design of this systematic mapping study. Section IV presents the results of the study and Section V discusses these results highlighting the main findings of the study. In Section VI, we discuss the threats to the validity of the study while we discuss the related work in Section VII. Section VIII concludes this research paper.

## II. BACKGROUND

This section presents the concepts related to gamification and its evaluation. Related works are further discussed in section VII.

Gamification is the use of mechanics from game domain in non-game contexts [1, 8]. The application of the mechanics defines goals and rules that drive users actions, allowing the emergence of dynamics, which are dependent on the users behavior. This organization leads to a "gamified experience" [8]. From the actions and reactions of users, emotions are evoked, which can keep users engaged and motivated to continue "playing" and, consequently, invested in the activities of the non-game context where the gamification was applied.

Hernández et al. [9] consider gamification a viable tool for software engineering education. It encourages students to effectively participate and interact in the learning process. For software engineering education, gamification is used, among other factors, to: improve social interaction; increase knowledge sharing; motivate the development of soft skills; improve team performance; and increase software quality. However, this broad range of applications comes at the cost of the complexity to adapt gamification for specific steps of software process [10]. In addition to that, tailoring the gamification approach requires significant effort from instructors [11].

Gasca-Hutardo et. al. [6] discuss that, despite the growing appearance of new gamification models, there has been few proposals for monitoring the designed gamification strategies. Klock et. al. [12] go further stating that the evaluation of gamification has not received the same level of attention that has been driven to search of new ways to apply gamification. Klock et al. [12] analyze 179 gamification studies in education, out of which only 20 evaluates the gamification design.

## III. STUDY DESIGN

This section describes the goal, research questions, and methods for the execution of this study.

### A. Goal and Research Question

The goal of this study is to identify and analyze the evaluation strategies used in the gamification literature, specifically in the context of its adoption to support software engineering practice and education. We are interested in understanding what goals, approaches, criteria, and data analysis procedures are used when evaluating the adoption and results of gamification. Therefore, we propose three research questions:

- *RQ1. Do the gamification studies in software engineering describe evaluation strategies?* – The goal is to map studies that propose or adopt any evaluation strategy,

- *RQ2. What is the purpose of the evaluation in gamification studies in software engineering?* – The goal is to classify the studies identified in RQ1 regarding the focus of the proposed or performed evaluation strategy,

- *RQ3. How is the evaluation performed in gamification studies in software engineering?* – The goal is to map the criteria, the type of approach (qualitative or quantitative), and data analysis procedure used in the evaluation strategies identified in RQ2.

### B. Method

To achieve the goal of this study, we performed a Systematic Mapping Study (SMS). SMS is a secondary study method that systematically (i.e., based on a structured and repeatable process or protocol) explores and categorizes studies in a given research field, and provides a structure of the type of research reports and results that have been published [13]. The selection of this research method is based on the nature of our research questions, which are based in the identification, classification, and quantification of evaluation strategies in the context of gamification studies in software engineering literature.

We executed the SMS in the period of April/2020 to August/2020. The study was organized in four steps, adapted from Petersen et al. [13], as follows:

- Step 1 – Definition of research questions: We have defined three research questions based on the study goal (Section III.A),

- Step 2 – Search: Based on the research questions, we defined and performed a replicable process for searching for retrieving papers in selected scientific databases (Section III.C),

- Step 3 – Study selection: We have defined and applied a replicable process for selecting only the relevant papers to address the goal of this study (Section III.D),

- Step 4 – Study Classification and Data Extraction: Based on the research questions, we have defined a strategy for both: (i) mapping the relevant data from the primary studies (Section III.E), and (ii) presenting the study results (Section IV).

Five researchers participated in the planning and execution of the study: an undergraduate and a graduate student in Computer Science, and three PhD software engineering professors/researchers.

### C. Search Strategy

In order to identify the primary studies for this SMS, the study considered the studies of Souza et al. [3] and Pedreira et al. [14]. The first mapped game-related approaches in Software Engineering education, which included the use of gamification, serious games, and game development activities. The latter mapped gamification studies in the context of software engineering practice. However, these studies only mapped primary studies up to 2016. Therefore, we extended the search using the same search string of Souza et al. [3] to collect studies up to the first quarter of 2020. The search string adopted included terms related to three domains: (a) education; (b) software engineering; and (c) game-related approaches. The terms searched were:

- Education: teach, learn, education, train,
- Software engineering: software engineering, software process, soft-ware requirements, requirements engineering, software verification, software validation, software design, software architecture, software test, software quality, software project management,
- Game-related approaches: game, serious games, edutainment, Gamification, game-based learning, gbl.

It is important to notice that this study is a subproduct of larger study. Therefore, the search string has terms related to other game-related approaches (game-based learning and game development based learning). Nonetheless, our inclusion and exclusion criteria considered only primary studies related to gamification.

The search string was applied in IEEE Xplore and ACM DL bases. We did not conduct search in EI COMPENDEX and SCOPUS because, according to the results of Souza et al. [3], there was a high rate of redundancy.

*D. Study Selection*

In this step we filtered only the relevant the primary studies to address our research questions. The inclusion and exclusion criteria defined and applied by three researchers are presented as follows:

- Exclusion criteria: Papers not written in English; papers not fully available to download; studies formatted as short papers (2 or less pages), tutorials, demos, books, book chapters, and similar; secondary studies; and duplicated studies,
- Inclusion criteria: studies whose main focus was on proposal, usage, discussion or evaluation of gamification in the context of software engineering education or practice,

The studies obtained were subject of a selection process in four stages: (i) three researchers screened the titles and abstract of all studies applying exclusion criteria; (ii) the researchers discussed over divergences on the application of exclusion criteria to reach consensus; (iii) the researchers read the abstract, and full text if necessary, to apply the inclusion criteria; (iv) the researchers discussed over divergences on the application of exclusion criteria to reach consensus. This selection process resulted in 100 primary studies (available in https://doi.org/10.5281/zenodo.4460065).

*E. Study Classification and Data Extraction*

To collect the data required for addressing the proposed research questions, three researchers were responsible for reading the 100 studies.

The data analysis consists in the classification of the studies in conformance to the proposed research questions. Therefore, the result of this SMS should map and classify studies regarding: the presence of evaluation strategies for gamification; the context in which gamification was used; number of participants in the evaluation; the goal of the evaluation; the evaluation criteria; the procedures to collect and analyze data related to the evaluation criteria.

IV. RESULTS

This section presents the results of the SMS. Subsection A presents an overview of the results. Subsection B, C and D describes the results for RQ1, RQ2, and RQ3, respectively. In these subsections, the codes will be used to reference the primary studies and is available at the previously defined URL.

*A. Overview*

The SMS resulted in the analysis of 100 primary studies, published between 2011 and 2020. Fig. 1 presents a histogram with the frequency of studies per year. There is clearly a growing tendency in the number of gamification studies over the years, with a positive spike from 2017 on. It illustrates the research community continued interest on the topic over the 10 years under analysis. The number of studies in 2020 is lower as a consequence of the period of execution of this study.

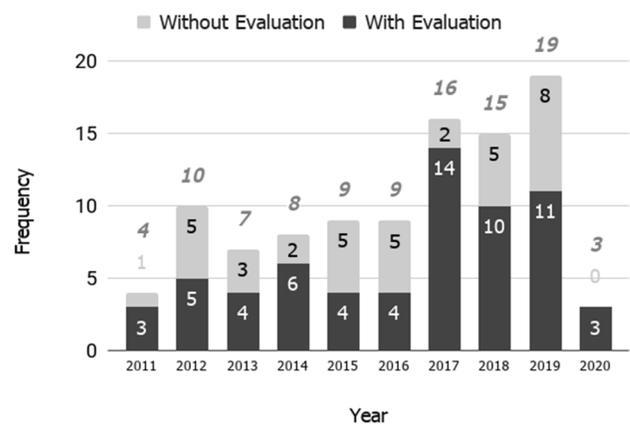

Fig. 1. Timeline of primary studies highlighting studies with and without evaluation strategies. (N=100)

Fig. 2 illustrates the distribution of primary studies regarding the context of application for gamification. It includes: software engineering teaching and assessment; software development; computer science teaching and assessment; civic participation; industrial production; health care; and general. The latter category is used for studies that propose a generic model or framework for gamification that can be applied in different contexts [PS040, PS047, PS081]. Each category is segmented using colors to denote studies that report procedures for the evaluation of gamification and studies without evaluation.

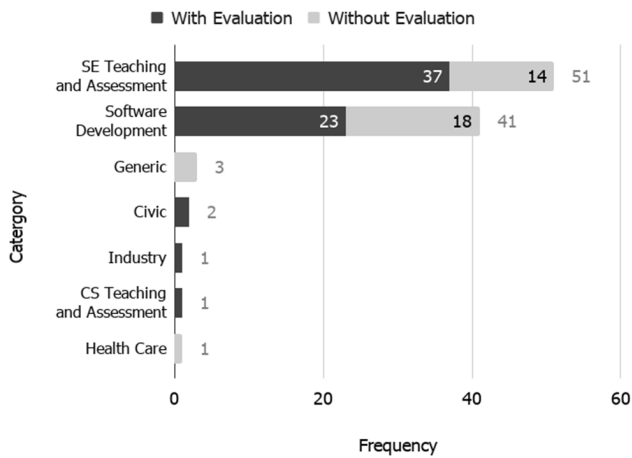

Fig. 2. Frequency of studies for each context of application of gamification approaches (N=100).

*B. Evaluating gamification*

This subsection presents results addressing RQ1 ("*Do the gamification studies in software engineering describe evaluation strategies?*").

Sixty-four (out of 100) primary studies presented some kind of evaluation strategy for gamification (see Table I). This includes both the evaluation of the application of gamification described in the studies and the proposal of evaluation models for gamification.

Fig. 1 uses colors to indicate the proportion of studies that describes (or not) evaluation strategies. In all years, the proportion of studies that describes an evaluation approach is higher or equal (only in 2012) than the studies that do not.

TABLE I. LIST OF PRIMARY STUDIES WITH AND WITHOUT EVALUATION

| Primary Studies with evaluation | Primary Studies without evaluation |
|---|---|
| [PS001], [PS002], [PS003], [PS004], [PS005], [PS006], [PS007], [PS012], [PS013], [PS015], [PS016], [PS017], [PS018], [PS021], [PS022], [PS023], [PS025], [PS027], [PS029], [PS031], [PS032], [PS034], [PS037], [PS038], [PS039], [PS041], [PS044], [PS045], [PS046], [PS048], [PS049], [PS050], [PS053], [PS056], [PS057], [PS058][PS059], [PS060], [PS061], [PS063], [PS066], [PS067], [PS068], [PS069], [PS072], [PS073], [PS075], [PS076], [PS078], [PS080], [PS083], [PS084], [PS086], [PS087], [PS088], [PS090], [PS091], [PS092], [PS093], [PS094], [PS096], [PS097], [PS098], [PS099] | [PS008], [PS009], [PS010], [PS011], [PS014], [PS019], [PS020], [PS024], [PS026], [PS028], [PS030], [PS033], [PS035], [PS036], [PS040], [PS042], [PS043], [PS047], [PS051], [PS052], [PS054], [PS055], [PS062], [PS064], [PS065], [PS070], [PS071], [PS074], [PS077], [PS079],[PS081], [PS082], [PS085], [PS089], [PS095], [PS100] |

Only three studies specifically propose evaluation models for gamification [PS046, PS078, PS097]. Sasso et al. [PS046] propose a model based in 5 dimensions: success metrics, analytics, conflicts, Jen Ratio[1], and Survey. For each dimension users should define goals, metrics, instruments, and presentation format for results.

Gasca-Hurtado et al. [PS078] present a layered framework with steps for the gamification of software processes. One of the layers is specifically designed towards an evaluation approach. The authors propose the evaluation method should be tailored for specific contexts considering basic principles (i) for the evaluation, (ii) for the procedures, and (iii) for the results. Ren et al. [PS097] also describe a layered model, organized in steps that range from the design to the evaluation of a gamification approach. The evaluation consists in the selection of metrics and adequate measurement methods in anticipation for the implementation of the gamification approach.

Only one of these studies actually applies their models in a case study [PS078]. However, the study does not report the execution of the evaluation layer. It was justified by the unavailability of required tools for the evaluation and by the development state of the model.

The remaining 61 primary studies applied specific strategies to evaluate their applications of gamification. For instance, Chow and Huang [PS001] used an experiment to evaluate the gamification framework proposed in their study. The author used a control and an intervention group to compare their performances and efficiency in a cross-cultural environment. The evaluation study produced objective and subjective data, which were analyzed quantitatively. The evaluation considered metrics such as Number of Lines of Code (LOC), covering rate of unit tests, and subjective factors related to performance such as the team's self-assessment using Likert scale questions.

As for the distribution of population sampling size in the primary 61 studies, we did not consider the 3 primary studies [PS046, PS078, PS097] that described evaluation models, because they did not report the execution of evaluation. Additionally, seven studies [PS007, PS012, PS016, PS018, PS056, PS061, PS075] did not report the number of participants in the evaluation procedures executed. The evaluation studies in the remaining 54 primary studies presented an average of 114,9 participants, with a median value of 28 and standard deviation of 466,32. There are seven outliers' studies [PS027, PS029, PS032, PS039, PS049, PS083, PS089] with over 131 participants.

*C. Goals of gamification evaluation*

This subsection presents results to address RQ2 ("*What is the purpose of the evaluation in gamification studies in software engineering?*").

From the analysis of the 64 primary studies discussed in subsection B, we observed two categories of evaluation approaches with respect to their purpose: evaluation of the gamification approach and evaluation of the gamification outcomes. The first is related to the evaluation of the users experience and opinions toward the gamification. The latter is related to the evaluation of the impact of gamification on the users, regarding the purpose of the gamification strategy implemented.

Fig. 3 and Fig. 4 show the distribution of primary studies regarding the purpose of their evaluation strategies. Most of the studies (44) evaluate both dimensions, the gamification approach itself and its outcomes. Only 20 studies focus on the evaluation of a single dimension: 11 studies evaluate only the gamification strategy and 9 studies evaluate only the outcomes of gamification. Fig. 4 shows the focus on which the assessment takes place (SE - Software Engineering, CS - Computer Science).

---

[1] Ratio between the total amount of positive interactions and negative interactions among users within a period of time.

For instance, Nikkila et. al. [PS021] consider five criteria in the evaluation of their approach: "engagement", "usability", "fun", "satisfaction", and "awareness". "Usability", "fun", and "satisfaction" are criteria related the assessment of the user experience while participating in the gamification execution. The criteria "engagement" and "awareness" are used for the assessment of the number of tasks finished and the level of awareness of the developer regarding the work done by the participant and other collaborators, respectively. Therefore, these two criteria are related to the evaluation of the outcomes of the gamification approach, and its effectiveness towards its purpose.

*1) Evaluation Criteria:* We identified 21 evaluation criteria used in the primary studies. The top 4 criteria were "engagement" (30 studies), "performance" (29 studies), "satisfaction" (28 studies), and "motivation" (19 studies). Other recurrent criteria found are: "usability" (9 studies), "learning" (6 studies), "utility" (5 studies), "efficiency" (5 studies), "perception" (4 studies), and "fun" (3 studies). The least used criteria (used in up to 2 studies) are: "communication", "user experience", "acceptance", "involvement", "sociability", "kindness", "confidence", "flow", "gameplay", "awareness" and "consciousness". The criteria frequency is presented in pairs of figures, segmented by the purpose of the evaluation strategy (gamified strategy vs effects of gamification), and grouped by: context (Fig. 5 and Fig. 6), data type (Fig. 7 and Fig. 8), and data analysis approach (Fig. 9 and Fig. 10). Fig. 5 and Fig. 6 present the frequency of studies that used each criterion for the evaluation of the gamification strategy and for the evaluation of the gamification outcomes, respectively.

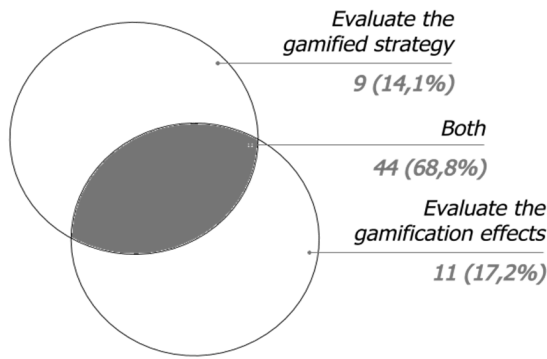

Fig. 3. Distribution according to the purpose of the evaluation strategy

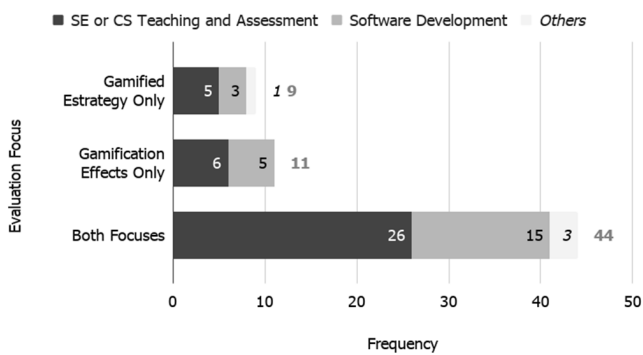

Fig. 4. Focus of the gamification evaluation (N=64)

*D. Methods for evaluating gamification*

This subsection presents results to address RQ3 ("*How is the evaluation performed in gamification studies in software engineering?*"). To address this research question we analyzed the primary studies regarding (i) the criteria, (ii) the type of data (objective and/or subjective), and (iii) the data analysis strategy used in the evaluation described in each primary study.

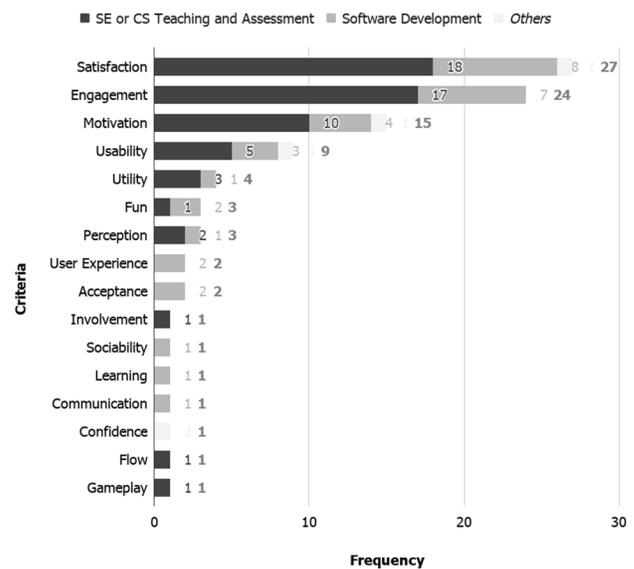

Fig. 5. Criteria for the evaluation of the gamification strategy, segmented by context of application (N=53).

For the evaluation of the gamification strategy, the most recurring criteria are satisfaction (27 studies), engagement (24 studies), and motivation (15 studies), as presented in Fig. 5. For the evaluation of the gamification outcomes, the most used criteria are performance, engagement, and motivation, as presented in Fig. 6. Therefore, engagement and motivation are criteria that are largely used for both purposes.

Additionally, we observed that some criteria are used exclusively for each purpose. "Performance", "efficiency", "awareness", and "kindness" are used exclusively in the evaluation of the gamification outcomes. "Fun", "acceptance", "sociability", "confidence", "flow" and "gameplay" are used exclusively for the evaluation of the gamification strategy. Considering the contexts of education and practice, the frequency distribution of the criteria is similar. The criteria "Learning", "Involvement", "Flow" and "Gameplay" are exclusively found in studies related to the software engineering education context. The criteria "User experience", "Acceptance", "Sociability", "Communication", "Kindness" "Awareness" and "Consciousness" are exclusively found in studies related to the software engineering practice context.

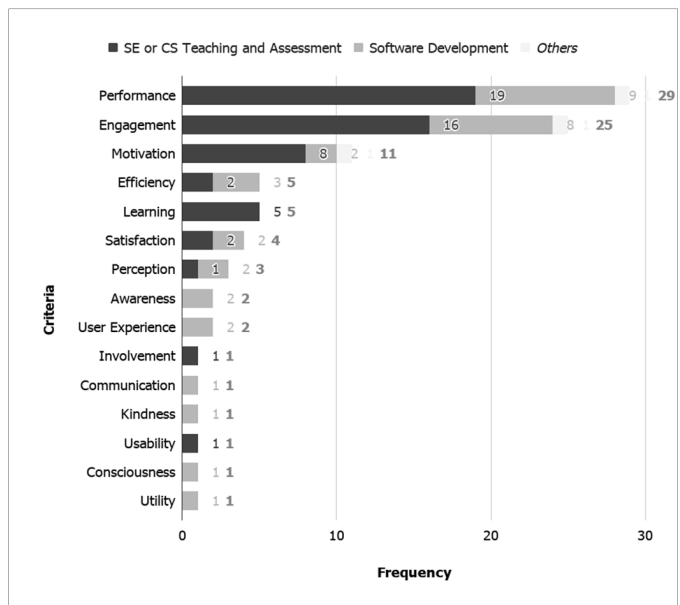

Fig. 6. Criteria for the evaluation of the gamification outcomes, segmented by context of application (N=55).

*2) Evaluation Approach:* The studies collect data for each criterion in an objective and/or subjective approach. Fig. 7 and Fig. 8 show the distribution of these data collection approaches for each category of evaluation.

The evaluation of the gamification strategy is mostly subjective (Fig. 7). This evaluation is targeted at the user opinion regarding their experience in the gamification. Therefore, this metrics are usually collected through opinion survey studies, with questionnaires and interviews. However, metrics for the evaluation of the "engagement" and "sociability" criteria are also collected objectively. For instance, Stanculescu [PS039] uses both Likert scale questionnaires and objective metrics such as session time and life span.

For the evaluation of the gamification outcomes (Fig. 8), the studies use both objective and subjective data. Metrics for "performance", "engagement", and "efficiency" are mostly objective data, while "learning" uses both strategies proportionally. For instance, instruments such as logs of operation and software for code review might be used to collect such data.

Tables II, III, IV, V and VI present the most used instruments to collect metrics for the analysis of the "satisfaction", "engagement", "motivation", "usability" and "performance" criteria, respectively. Questionnaires, interviews, observation, professor evaluation and logs are the most used instruments for data collection.

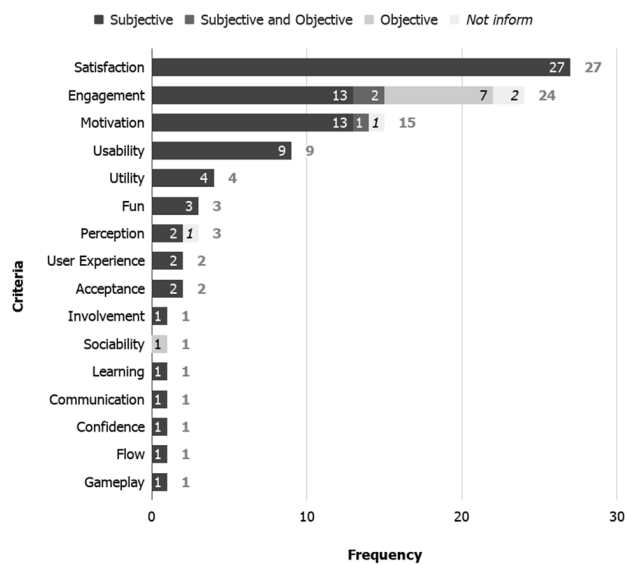

Fig. 7. Criteria for the evaluation of the gamification strategy, segmented by data type (N=53).

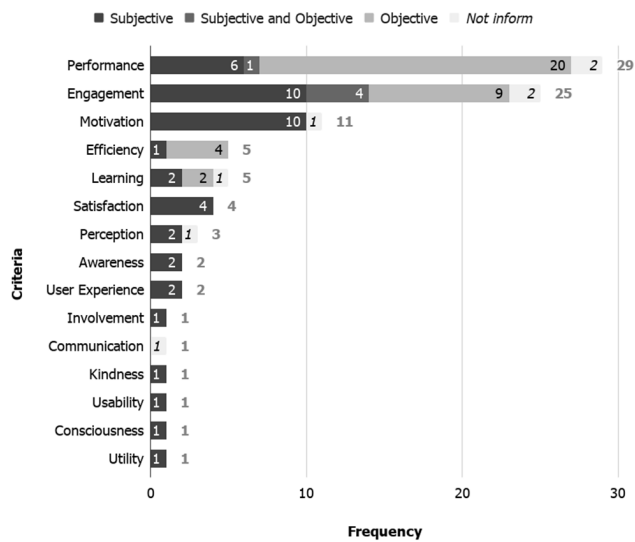

Fig. 8. Criteria for the evaluation of the gamification outcomes, segmented by data type (N=55).

TABLE II. DATA COLLECTION INSTRUMENTS FOR THE "SATISFACTION" CRITERION.

| Focus | Instrument | Primary Studies | # |
|---|---|---|---|
| Gamification Strategy | Questionnaire | [PS002], [PS004], [PS005], [PS013], [PS021], [PS023], [PS027], [PS038], [PS041], [PS059], [PS066], [PS067], [PS073], [PS075], [PS080], [PS083], [PS084], [PS091], [PS092], [PS099] | 20 |
| | Interview | [PS013], [PS017], [PS023], [PS027], [PS053], [PS059], [PS073] | 7 |
| | Observation | [PS007], [PS021], [PS091] | 3 |
| | Survey in forums and e-mail lists | [PS046] | 1 |
| | Consulting with specialists in gamification | [PS060] | 1 |
| | Neurocomputing | [PS078] | 1 |
| | Not reported | [PS022] | 1 |

TABLE III. DATA COLLECTION INSTRUMENTS FOR THE "ENGAGEMENT" CRITERION.

| Focus | Instrument | Primary Studies | # |
|---|---|---|---|
| Gamification Strategy | Questionnaire | [PS020], [PS003], [PS004], [PS031], [PS038], [PS039], [PS053], [PS060], [PS066], [PS080] | 10 |
| | Operation logs | [PS013], [PS018], [PS031], [PS032], [PS039], [PS046], [PS049], [PS057] | 8 |
| | Interview | [PS022], [PS023], [PS025], [PS090] | 4 |
| | Observation | [PS07], [PS080] | 2 |
| | Student presence | [PS090] | 1 |
| | Not reported | [PS075], [PS087] | 2 |
| Gamification outcomes | Operation | [PS013], [PS015], [PS018], [PS021], [PS029], [PS031], [PS032], [PS037], [PS046], [PS049], [PS057], [PS096] | 12 |
| | Questionnaire | [PS002], [PS003], [PS004], [PS015], [PS031], [PS038], [PS060], [PS066], [PS080] | 9 |
| | Interview | [PS025], [PS037], [PS090], [PS096] | 4 |
| | Observation | [PS007], [PS080] | 2 |
| | Student presence | [PS006], [PS090] | 2 |
| | Tasks performed | [PS006] | 1 |
| | Not reported | [PS075], [PS087] | 2 |

TABLE IV. DATA COLLECTION INSTRUMENTS FOR THE "MOTIVATION" CRITERION.

| Focus | Instrument | Primary Studies | # |
|---|---|---|---|
| Gamification Strategy | Questionnaire | [PS003], [PS013], [PS038], [PS050], [PS059], [PS060], [PS063], [PS068], [PS069], [PS084] | 10 |
| | Interview | [PS015], [PS022], [PS037], [PS059] | 4 |
| | Not reported | [PS075] | 1 |
| Gamification outcomes | Questionnaire | [PS003], [PS013], [PS017], [PS027], [PS056], [PS060], [PS068], [PS086], [PS093] | 9 |
| | Interview | [PS027], [PS037] | 2 |
| | Not reported | [PS075] | 1 |

TABLE V. DATA COLLECTION INSTRUMENTS FOR THE "USABILITY" CRITERION.

| Focus | Instrument | Primary Studies | # |
|---|---|---|---|
| Gamification Strategy | Questionnaire | [PS004], [PS021], [PS044], [PS048], [PS060], [PS066], [PS072], [PS076], [PS098] | 9 |
| | Interview | [PS098] | 1 |

TABLE VI. DATA COLLECTION INSTRUMENTS FOR THE "PERFORMANCE" CRITERION.

| Focus | Instrument | Primary Studies | # |
|---|---|---|---|
| Gamification outcomes | Professor evaluation | [PS02], [PS03], [PS06], [PS07], [PS016], [PS032], [PS083], [PS086], [PS088], [PS091], [PS092], [PS093] | 12 |
| | Questionnaire | [PS001], [PS013], [PS027], [PS038], [PS044], [PS050], [PS056], [PS061] | 8 |
| | Code review software | [PS001], [PS002], [PS012], [PS016], [PS023] | 5 |
| | Performance parameters | [PS041], [PS067], [PS099] | 3 |
| | Operation logs | [PS032], [PS096], [PS099] | 3 |
| | Not reported | [PS034], [PS087], [PS097] | 3 |

*3) Data Analysis:* Fig. 9 and Fig. 10 categorize the data analysis methods used for each criterion as qualitative and/or quantitative, for the evaluation of the gamification strategy and evaluation of the gamification outcomes, respectively.

The data analysis for the assessment of criteria related to the evaluation of gamification outcomes assumes a predominant quantitative nature. Yet, there are still qualitative methods present.

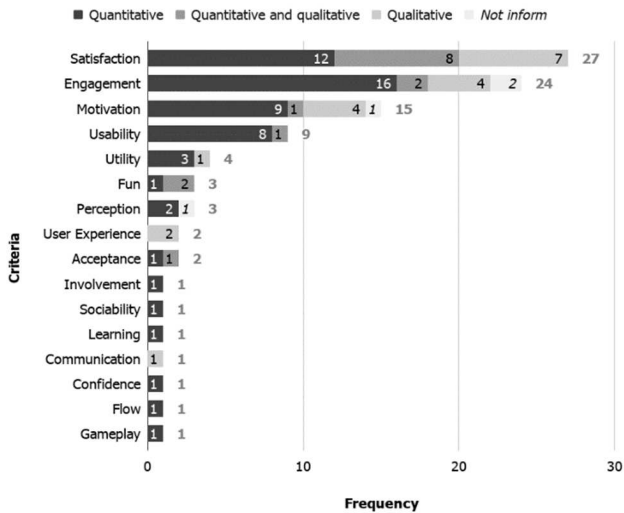

Fig. 9. Criteria for the evaluation of the gamification strategy, segmented by data analysis approach (N=53).

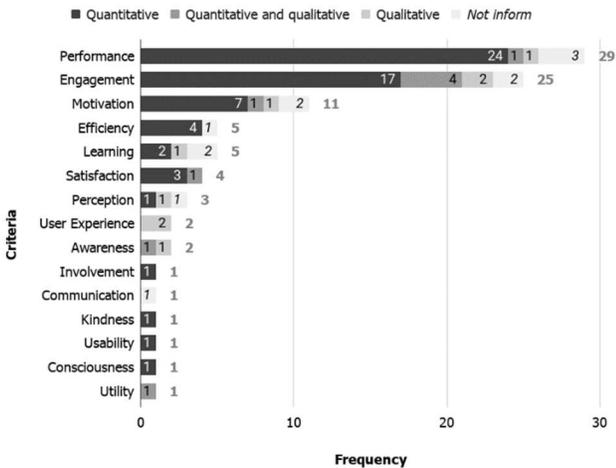

Fig. 10. Criteria for the evaluation of the gamification outcomes, segmented by data analysis approach (N=55).

## V. Discussion

This section presents our main findings and impressions on the results presented in Section IV.

Our main finding is the low number of studies proposing models for the evaluation of gamification. We only found three studies proposing this type of models, yet we did not find any evidence supporting the adoption of these methods. The immediate consequence is the lack of standardization in the evaluation of applications of gamification, both regarding methods and criteria. An implication is that, in the primary studies analyzed, there were no cases of studies comparing their proposals with others. Considering Game-based learning approaches in software engineering there are studies that propose standard evaluation methods and instruments to allow the assessment and comparison of serious games, such as MEEGA+ [15].

We acknowledge that the range of applications for gamification is wide, even if we consider only the scope of software engineering. Yet, our second finding is that the primary studies focus on two aspects when evaluating gamification application: the (i) gamification strategy itself, and the user experience and impressions; and the (ii) outcomes and effects of gamification. The first is more prone to be the target of a generic evaluation model, as its focus is assessing the level of engagement, satisfaction, motivation, fun, and other feelings and perceptions that rise from the user interaction with the gamification application. The latter is more context-sensible, as different applications of gamification aims at different outcomes, effects and impacts on users and on the non-game context. However, it is possible to suggest guidelines to support gamification designers to clearly state the goals of gamification and useful criteria that can be tailored for each specific context (e.g. "performance").

Our third finding is the recurring set of criteria used for the evaluation of gamification: "engagement", "motivation", "satisfaction", and "performance". First, it is interesting to notice that "fun" is not present, in contrast to a possible erroneous understanding that gamification is meant to make activities fun. As one can conclude from our results, the goal of gamification is to engage and motivate users in activities of a non-game context, which usually aims at improving their performance. In order to achieve such goals, the gamification strategy itself should sustain the users' feelings of satisfaction, motivation, and engagement. Therefore, "engagement" and "motivation" are both a mean and an end. That is, in order to keep users engaged and motivated in the non-game context, users must also be engaged and motivated in the mechanics and dynamics of the gamification application. Alternatively, "performance" is strictly dependent on the non-game context. For software engineering education performance may be related to the quality, number or frequency of artifacts produced by learners, while in software development it may be related to productivity or quality. However, performance is often a quantitative and objective metric to assess the impact of gamification on the activities of the users in the non-game context.

Finally, our fourth finding is that the evaluation of gamification must consider both subjective and objective data. Gamification is closely related to evoking users' feelings and addressing intrinsic and extrinsic motivations. Therefore, the evaluation of the gamification strategy needs subjective inputs from users' impressions and opinions. However, the evaluation of gamification outcomes has ample room for the collection of objective data, especially when we consider criteria such as "performance". Yet, neither of these aspects are strictly subjective nor objective. For instance, in software engineering education, measuring improvements on learning is not an easy task, and students' self-assessment input is a subjective data that can be used in conjunction with grades.

### A. Evaluation of gamification in Software engineering education

Our primary study sample has 52 studies that uses gamification in the context of software engineering education. From these, 37 primary studies reported evaluation studies. Regarding the focus of the evaluation, 27 primary studies ($\cong$71.05%) evaluated both the strategy and the outcomes of gamification, 6 ($\cong$15.79%) focused only on the evaluation of the strategy, and 7 (≅18.42%) only on the outcomes. The reason for this is that these primary studies had the main focus on validating the gamification approach with learners and lecturers [16, 17, 18]. Only two primary studies compared the gamification approach with other educational methods [19, 20].

We did not find studies that proposed gamification evaluation methods specifically for the context of software engineering education. The goal of all studies was to report and/or evaluate the results of introducing specific applications of gamification in software engineering education.

It was surprising that only 5 primary studies used "learning" as an evaluation criterion. We observed that "performance", "engagement", and "motivation" are often used as criteria to evaluate the expected outcomes of gamification. This result reinforces the auxiliary role of gamification, as an approach to support lecturers in sustaining students' participation and interest during the learning process, rather than targeting directly at improving learning.

Jiang [4] states the scarcity of studies assessing the impact of gamification elements in software engineering education. Our sample included 7 studies with this type of analysis [11, 21, 22, 23, 24, 25, 26]. Therefore, our results support Jiang [4]. More empirical data is necessary to understand the relation between specific gamification elements and their effects on learning.

## VI. THREATS TO VALIDITY

This section discusses the possible threats to the validity of this study and the actions taken to address validity issues. We used the structure proposed by Wohlin et al. [27].

### A. Construct Validity

To minimize the risk that study was not able to address the research questions, a pilot study was performed using the primary studies obtained from published SMS results from Souza et al. [3] and Pedreira et al. [14]. Only after we validated our research questions we expanded the search for additional primary studies.

### B. Internal validity

During the extraction process, the studies were classified based on our judgment. However, despite double-checking, some studies could have been classified incorrectly, especially regarding the mapping of the evaluation criteria, which we not always explicitly reported in the primary studies.

### C. External validity

It is possible that the systematic mapping study did not return all relevant studies on gamification of software engineering and evaluation models for this context. To mitigate this risk, we chose to expand previous secondary studies [3, 14], rather than initiating a new one from the scratch.

### D. Conclusion validity

To ensure conclusion validity of our study, we presented throughout Section IV graphs and tables exposing results generated directly from the data, and discussed the explicit observations and trends. This ensures a high degree of traceability between the data and conclusions. In addition, our corpus of studies is available for other researchers. In addition to that, the SMS process was executed with the support of three authors with previous experience and publications in gamification and software engineering.

## VII. RELATED WORK

This section presents similar studies that are directly or indirectly related to the investigation of the present study. To the best of our effort, we did not find secondary studies that surveyed the literature for methods for the evaluation of gamification in software engineering. However, there are secondary studies that explore other aspects of gamification in software engineering [3, 14]

Souza et al. [3] conducted a SMS to map game-related methods for software engineering education. The study did not have the specific focus on gamification; however, it identified and mapped 10 gamification studies. The study did not provide in-depth analysis on evaluation strategies. The authors briefly commented that 70 (out of 156) primary studies formally describe proper evaluation methods and results, regarding effectiveness in learning or perception of users. From these only 6 (out of 10) gamification studies described evaluation methods. The present studies used the primary studies related to gamification identified by Souza et al. [3] and its search string to expand its results.

Similarly, Pedreira et al. [14] identified 29 primary studies describing the use of gamification in the professional context of software engineering. The authors state that only few primary studies offer empirical evidence on the impact of their proposals on user engagement and performance. The authors point the necessity of further research providing empirical results about the effect of Gamification. We also considered the primary studies of Pedreira et al. [14] for the initial sample of primary studies, expanding their results and focusing on the evaluation aspect, specifically.

We found three secondary studies that investigate the effects of gamification in education for the evaluation of gamification [4, 28, 29]. However, none of these studies investigate the specific context of software engineering.

Jiang [4] conducted a systematic review to investigate the effectiveness of implementing gamification in education and to identify the factors influencing its effectiveness. The authors selected 18 empirical studies, from 2013 to 2016, on experiments and quasi-experiments on the application of gamification in education. However, the author did not investigate how the evaluation of gamification was executed in each primary study. Nevertheless, Jiang [4] points to the limitations of the research design in these empirical studies, which may interfere in their conclusion validity. The author also reinforces the position about the need of additional studies to understand the effects of gamification.

Bai, Hew e Huang [28] conducted a meta-analysis of 30 independent interventions (3,202 participants) drawn from 24 quantitative studies that have examined the effects of gamification on student academic performance in various educational settings.

Finally, Klock et al. [29] conducted a SMS to investigate

how studies applying gamification evaluate this technique in educational environments. The authors selected 20 primary studies, from 2013 e 2016. The authors found that these studies compared and evaluated gamification through students' interaction, performance, and user experience, based on their activities and answers in satisfaction surveys and tests.

Although these studies [4, 28, 29] are not focused on software engineering context, we believe that their results are converging with ours. Our study complements the results of Klock et al. [29], by expanding the analysis to a significantly larger sample of primary studies, while focusing on the context of software engineering. However, Klock et al. [29] go further analyzing the game elements present in each study and correlate the results of the evaluation of each study with these elements.

## VIII. CONCLUSION

This study described a SMS investigating strategies for the evaluation of gamification in software engineering. We selected 100 primary studies, from 2011 to 2020. From these, 64 studies report procedures for the evaluation of gamification. However, only three studies actually propose evaluation models for gamification. We observed that the evaluation of gamification focuses on two aspects: the evaluation of the gamification strategy itself, related to the user experience and perceptions; and the evaluation of the outcomes and effects of gamification on its users and context. The most recurring criteria for the evaluation are "engagement", "performance", "satisfaction", and "motivation". Finally, the evaluation of gamification requires a mix of subjective and objective inputs, and qualitative and quantitative data analysis approaches. Depending of the focus of the evaluation (the strategy or the outcomes), there is a predominance of a type of data and analysis.

From our results, we observed the importance of standardization in the design and report of the evaluation of gamification. It is not viable to compare gamification applications from the primary studies, given the diversity of methods. Despite the identification of three evaluation models, we did not find any study that actually adopts them. Therefore, we believe that the community should strive for standard procedures to allow (at least some level of) comparison of applications of gamification, in order to really generate meaningful empirical data for the understanding of the impacts of gamification in different contexts, setups and users.

For future work, we plan to define guidelines for the evaluation of gamification, based on the results of this SMS.

## IX. ACKNOWLEDGEMENTS

This work was supported by CAPES (grant 88887.485345/2020-00).